\documentclass[preprint,floats,epsf,epsfig,nofootinbib]{revtex4}
\usepackage{epsfig}

\begin{document}
\def\be{\begin{eqnarray}}
\def\en{\end{eqnarray}}
\def\non{\nonumber}
\def\la{\langle}
\def\ra{\rangle}
\def\nc{N_c^{\rm eff}}
\def\vp{\varepsilon}
\def\drho{\bar\rho}
\def\deta{\bar\eta}
\def\CP{{\it CP}~}
\def\a{{\cal A}}
\def\B{{\cal B}}
\def\c{{\cal C}}
\def\d{{\cal D}}
\def\e{{\cal E}}
\def\p{{\cal P}}
\def\t{{\cal T}}
\def\up{\uparrow}
\def\dw{\downarrow}
\def\vma{{_{V-A}}}
\def\vpa{{_{V+A}}}
\def\smp{{_{S-P}}}
\def\spp{{_{S+P}}}
\def\J{{J/\psi}}
\def\ov{\overline}
\def\Lqcd{{\Lambda_{\rm QCD}}}
\def\pr{{Phys. Rev.}~}
\def\prl{{Phys. Rev. Lett.}~}
\def\pl{{Phys. Lett.}~}
\def\np{{Nucl. Phys.}~}
\def\zp{{Z. Phys.}~}
\def\lsim{ {\ \lower-1.2pt\vbox{\hbox{\rlap{$<$}\lower5pt\vbox{\hbox{$\sim$}
}}}\ } }
\def\gsim{ {\ \lower-1.2pt\vbox{\hbox{\rlap{$>$}\lower5pt\vbox{\hbox{$\sim$}
}}}\ } }

\font\el=cmbx10 scaled \magstep2{\obeylines\hfill November, 2005}

\vskip 1.5 cm

\centerline{\large\bf Penguin-induced Radiative Baryonic $B$
Decays Revisited}

\bigskip
\centerline{\bf Hai-Yang Cheng$^1$ and Kwei-Chou Yang$^2$}
\medskip
\centerline{$^1$ Institute of Physics, Academia Sinica}
\centerline{Taipei, Taiwan 115, Republic of China}
\medskip
\centerline{$^2$ Department of Physics, Chung Yuan Christian
University} \centerline{Chung-Li, Taiwan 320, Republic of China}
\bigskip
\bigskip
\bigskip
\bigskip
\bigskip
\centerline{\bf Abstract}
\bigskip
\small

Weak radiative baryonic $B$ decays $\ov B\to\B_1\ov \B_2\gamma$
mediated by the electromagnetic penguin process $b\to s\gamma$ are
re-examined within the framework of the pole model. The meson pole
contribution that has been neglected before is taken into account
in this work. It is found that the intermediate $K^*$ contribution
dominates in the $\Sigma\bar p\gamma$ mode and is comparable to
the baryon pole effect in $\Lambda\bar p\gamma$ and
$\Xi\bar\Lambda\gamma$ modes. The branching ratios for
$B^-\to\Lambda\bar p\gamma$ and $B^-\to\Xi^0\bar\Sigma^-\gamma$
are of order $2.6\times 10^{-6}$ and $6\times 10^{-7}$,
respectively. The threshold enhancement effect in the dibaryon
mass spectrum is responsible by the meson pole diagram. We also
study the angular distribution of the baryon in the dibaryon rest
frame. The baryon pole diagrams imply that the antibaryon tends to
emerge in the direction of the photon in the baryon-pair rest
frame. The predicted angular asymmetry agrees with experiment for
$B^-\to\Lambda\bar p\gamma$. Measurements of the correlation of
the photon with the baryon allow us to discriminate between
different models for describing the radiative baryonic $B$ decays.
For decays $B\to\Xi\bar\Sigma\gamma$, a large correlation of the
photon to the $\bar\Sigma$ and a broad bump in the dibaryon mass
spectrum are predicted.

\pagebreak

{\bf 1.}~~In \cite{CYrad} we have studied weak radiative baryonic
$B$ decays  and pointed out that the decays $B\to\B_1\ov
\B_2\gamma$ mediated by the electromagnetic penguin process $b\to
s\gamma$ can have rates larger than their two-body counterparts
$B\to\B_1\ov \B_2$ due to the large short-distance enhancement for
the electromagnetic penguin transition. In particular, the
branching ratios for $B^-\to\Lambda\bar p\gamma$ and
$B^-\to\Xi^0\bar\Sigma^-\gamma$ are sizable, of order $1\times
10^{-6}$.\footnote{The prediction of $\B(B^-\to\Lambda\bar
p\gamma)$, for example, has been updated in \cite{Cheng03}.}
We thus concluded that the penguin-induced radiative baryonic $B$
decays should be accessible to $B$ factories. Recently, Belle
\cite{Belle:Lampgam} has made the first observation of the
radiative hyperonic $B$ decay $B^-\to\Lambda\bar p\gamma$ with the
result
 \be
 \B(B^-\to\Lambda\bar p\gamma)=(2.16^{+0.58}_{-0.53}\pm0.20)\times
 10^{-6}
 \en
and set an upper limit on the branching ratio of
$\B(B^-\to\Sigma^0\bar p\gamma)<4.6\times 10^{-6}$. Therefore, the
former radiative decay mode has a rate much larger than the
two-body counterpart, that is,  $\Gamma(B^-\to\Lambda\bar
p\gamma)\gg\Gamma(B^-\to\Lambda\bar p)$, where only the upper
limit of $4.9\times 10^{-7}$ has been set for the branching ratio
of the latter \cite{Belle:2body}.

In the past few years, there are two more theoretical studies on
radiative baryonic $B$ decays: one by Kohara \cite{Kohara} and the
other by Geng and Hsiao \cite{GH}. Based on the assumption of
penguin transition dominance and flavor SU(3) symmetry, Kohara has
derived several relations among the decay rates of various decay
modes. Geng and Hsiao have followed \cite{CHT02} to parametrize
the three-body $B$ to $\B_1\ov \B_2$ transition matrix element in
terms of several unknown form factors and then invoked the pQCD
counting rules to require the form factors to behavior as the
inverse powers of the dibaryon invariant mass squared
$M_{\B_1\ov\B_2}^2\equiv(p_{_{{\B_1}}}+p_{\bar\B_2})^2$. Although
their result on $\B(B^-\to\Lambda\bar p\gamma)$ is very similar to
ours, their predictions for some other radiative modes are quite
different from ours. For example, whereas $B^-\to\Sigma^0\bar
p\gamma$ is predicted to be quite suppressed, of order $10^{-9}$,
in our pole model calculation, it can be as large as $10^{-7}$ in
\cite{GH}.

In this work we would like to re-examine the radiative baryonic
$B$ decays within the pole model framework for two reasons. First,
in the previous work we have neglected the meson pole
contributions which are {\it a priori} not necessarily small. Also
some input parameters such as the strong couplings need to be
updated. Second, for the decay $B^-\to\Lambda\bar p\gamma$ Belle
has measured the angular distribution of the baryon in the
baryon-pair rest frame. It is found that the $\Lambda$ tends to
emerge opposite the direction of the photon. The correlation of
the photon with the $\Lambda$ is sensitive to the underlying decay
mechanism and hence can be used to discriminate between different
models.

\vskip 0.5cm {\bf 2.}~~As shown in \cite{CYrad}, the radiative
baryonic $B$ decays of interest that proceed through the
electromagnetic penguin mechanism $b\to s\gamma$ are
 \be \label{raddecays}
&& B^-\to\{\Lambda\bar p,\,\Sigma^0\bar
p,\,\Sigma^+\bar\Delta^{--},\,\Sigma^-\bar n,
 \,\Xi^0\bar\Sigma^-,\,\Xi^-\bar\Lambda,\,\Xi^-\bar\Sigma^0,\,\Omega^-\bar\Xi^0\}\gamma,
 \non \\ &&
 \ov B^0\to\{\Lambda\bar n,\,\Sigma^0\bar n,\,\Sigma^+\bar p,\,\Sigma^-\bar \Delta^+,
 \,\Xi^0\bar\Lambda,\,\Xi^0\bar\Sigma^0,\,\Xi^-\bar\Sigma^+,\,\Omega^-\bar\Xi^+\}\gamma.
 \en
Let us consider the decay $B^-\to\Lambda\bar p\gamma$ as an
illustration. The short-distance $b\to s\gamma$ penguin
contribution to $B^-\to\Lambda\bar p\gamma$ depicted in Fig. 1 is
described by the weak Hamiltonian
 \be
 {\cal H}_W=-{G_F\over\sqrt{2}}\,V_{ts}^*V_{tb}c_7^\gamma O_7,
 \en
with
 \be
 O_7={e\over 8\pi^2}\,m_b\bar
 s\sigma_{\mu\nu}F^{\mu\nu}(1+\gamma_5)b.
 \en
The Wilson coefficient $c_7^\gamma$ will be specified later. Since
a direct evaluation of this diagram is difficult as it involves an
unknown 3-body matrix element $M_{\mu\nu}=\la\Lambda\bar p|\bar
s\sigma_{\mu\nu}(1+\gamma_5)b|B^-\ra$, we shall instead evaluate
the corresponding diagrams known as pole diagrams at the hadron
level (see Fig. 1). In principle, there exist many possible baryon
and meson pole contributions. The main assumption of the pole
model is that the dominant contributions arise from the low-lying
baryon and meson intermediate states. For $B^-\to\Lambda\bar
p\gamma$, the relevant intermediate states are $\Lambda_b^{(*)}$
and $\Sigma_b^{0(*)}$.

\begin{figure}[tb]
\vspace{-1cm}
\hspace{0cm}\centerline{\psfig{figure=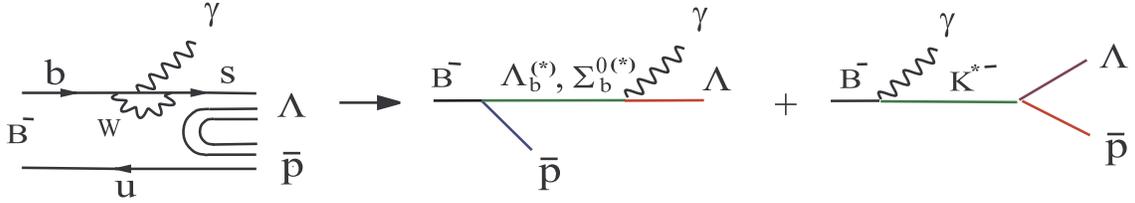,width=15cm}}
\vspace{0cm}
    \caption{{\small Quark and pole diagrams for $B^-\to\Lambda\bar
    p\gamma$.
    }}
\end{figure}

Let us consider the $\Lambda_b$ pole first. Since the intermediate
$\Lambda_b$ state amplitude
 \be
 A(B^-\to\Lambda\bar p\gamma(k))_{\Lambda_b}=g_{\Lambda_b\to B^-p}\la\Lambda
 \gamma|{\cal H}_W|\Lambda_b\ra\,{i\over (p_\Lambda+k)^2-m_{\Lambda_b}^2}
 \bar u_{\Lambda_b}\gamma_5 v_{\bar p}\,,
 \en
has been evaluated in \cite{CYrad}, we just write down its final
expression
 \be \label{eq:Lampole}
 A(B^-\to\Lambda\bar p\gamma)_{\Lambda_b}=-ig_{\Lambda_b\to B^-p}\,\bar
 u_\Lambda(a+b\gamma_5)\sigma_{\mu\nu}\vp^\mu k^\nu
 { p\!\!\!/_\Lambda+k\!\!\!/+m_{\Lambda_b}\over
 (p_\Lambda+k)^2-m_{\Lambda_b}^2}\gamma_5 v_{\bar p}\,,
 \en
with
 \be
 a &=& {G_F\over \sqrt{2}}\,{e\over
8\pi^2}2c_7^\gamma m_bV_{tb}V^*_{ts}\,[f_1^{\Lambda_b\Lambda}(0)
-f_2^{\Lambda_b\Lambda}(0)],   \non \\
b &=& {G_F\over \sqrt{2}}\,{e\over 8\pi^2}2c_7^\gamma m_bV_{tb}
V^*_{ts}\,\Big[g_1^{\Lambda_b\Lambda}(0)+{m_{\Lambda_b}-m_\Lambda\over
 m_{\Lambda_b}+m_\Lambda}g_2^{\Lambda_b\Lambda}(0)\Big],
 \en
where $f_i^{\Lambda_b\Lambda}$ and $g_i^{\Lambda_b\Lambda}$ are
the form factors for $\Lambda_b\to\Lambda$ transitions defined by
 \be
\la \B_1(p_1)|(V-A)_\mu|\B_2(p_2)\ra &=& \bar
u_1(p_1)\Bigg\{f_1^{\B_2\B_1}(q'^2)\gamma_\mu+i{f_2^{\B_2\B_1}(q'^2)\over
m_1+m_2} \sigma_{\mu\nu}q'^\nu+{f_3^{\B_2\B_1}(q'^2)\over
m_1+m_2}q'_\mu \non  \\ &-&
\Big[g_1^{\B_2\B_1}(q'^2)\gamma_\mu+i{g_2^{\B_2\B_1}(q'^2)\over
m_1+m_2} \sigma_{\mu\nu}q'^\nu+{g_3^{\B_2\B_1}(q'^2)\over
m_1+m_2}q'_\mu\Big]\gamma_5\Bigg\}u_2(p_2),
 \en
with $q'=p_2-p_1$. To obtain the result (\ref{eq:Lampole}) we have
applied the relation $\gamma_0 u_{\Lambda_b}=u_{\Lambda_b}$ valid
in the static heavy $b$-quark limit. As explained in \cite{CYrad},
the ${1\over 2}^-$ state $\Lambda_b^*$ and the pole states
$\Sigma_b^0$ and $\Sigma_b^{0*}$ do not contribute to
$B^-\to\Lambda\bar p\gamma$.

The $K^*$ meson pole contribution, which has been neglected in our
previous work, is {\it a priori} not necessarily suppressed. Its
amplitude reads
 \be \label{eq:Kpole}
  A(B^-\to\Lambda\bar p\gamma)_{K^*} = A(B^-\to K^{*-}\gamma)
 {i\over q^2-m_{K^*}^2}\,\bar u_{\Lambda}\,\vp^{\nu}_{K^*}i\left(h_1^{\Lambda
 pK^*}\gamma_\nu+i{h_2^{\Lambda pK^*}\over
 m_{\Lambda}+m_p}\sigma_{\nu\lambda}q^\lambda\right)v_{\bar p},
 \en
where the strong couplings $h_1$ and $h_2$ will be specified
shortly, $q=p_\Lambda+p_{\bar p}$, and
 \be
 A(B^-\to K^{*-}(q)\gamma(k)) &=&
 -{G_F\over\sqrt{2}}V_{ts}^*V_{tb}\,a_7^\gamma\,{e\over
 8\pi^2}\,2m_b\Bigg\{ i\epsilon_{\mu\nu\alpha\beta}\vp^\mu
 k^\nu\vp^\alpha_{K^*}q^\beta T_1(0) \non \\
 &+& \vp^\mu\left[\vp^\mu_{K^*}(m_B^2-m_{K^*}^2)-(p+q)_\mu\vp_{K^*}\cdot
 k\right]T_2(0)\Bigg\},
 \en
with $\vp$ and $\vp_{K^*}$ being the polarization vectors of the
photon and $K^*$, respectively. In the above expression, we have
replaced the Wilson coefficient $c_7^\gamma$ by the parameter
$a_7^\gamma$ to take into account vertex and spectator corrections
to $B\to K^*\gamma$. It is easily seen that the first term with
the form factor $T_1$ does not contribute to the $K^*$ pole
amplitude. Summing over the polarization of the intermediate $K^*$
state and applying the Gordon decomposition
 \be
 {i\over m_\Lambda+m_p}\bar u_\Lambda\sigma_{\mu\nu}q^\nu v_{\bar p}
=\bar u_\Lambda\gamma_\mu v_{\bar p}-{1\over m_\Lambda+m_p}\bar
u_\Lambda(p_\Lambda-p_{\bar p})_\mu v_{\bar p},
 \en
the $K^*$ pole amplitude becomes
 \be \label{eq:Kstamp}
  A(B^-\to\Lambda\bar p\gamma)_{K^*} &=& -{G_F\over\sqrt{2}}V_{ts}^*V_{tb}\,c_7^\gamma
\,{e\over 8\pi^2}\,T_2(0)\,\bar u_\Lambda\Bigg\{ {1\over
q^2-m_{K^*}^2}
[\vp^\mu(m_B^2-m_{K^*}^2)-2\vp\cdot p_Bk^\mu]  \\
&\times& \left[-(h_1^{\Lambda pK^*}+h_2^{\Lambda
pK^*})\gamma_\mu+h_1^{\Lambda pK^*}{m_\Lambda-m_p\over
m_{K^*}^2}(p_B)_\mu+{h_2^{\Lambda pK^*}\over
m_\Lambda+m_p}(p_\Lambda-p_{\bar p})_\mu\right]\Bigg\} v_{\bar p}.
\non
 \en
We shall follow \cite{CY03}  to relate the strong couplings $h_1$
and $h_2$ to the vector form factors defined in the matrix element
$\la \Lambda\bar p|(V-A)_\mu|0\ra$ via the intermediate $K^*$
pole; that is,
 \be
 h_1^{\Lambda pK^*}(q^2)=\,{q^2-m_{K^*}^2\over f_{K^*}m_{K^*}}f_1^{\Lambda p}(q^2),
 \qquad\quad
 h_2^{\Lambda pK^*}(q^2)=\,{q^2-m_{K^*}^2\over f_{K^*}m_{K^*}}f_2^{\Lambda
 p}(q^2).
 \en

There exist other strange kaon intermediate states such as the
excited $K^*$ ones and the $p$-wave states such as $K_0^*$, $K_1$
and $K_2^*$. However, they are expected to be suppressed at both
strong and weak vertices, namely, $\Gamma(B\to K_{\rm
resoance}\gamma)<\Gamma(B\to K^*\gamma)$ and $h^{\Lambda p K_{\rm
resonance}}<h^{\Lambda pK^*}$. Therefore, we will focus on the
dominant $K^*$ pole contribution in the present work.

To proceed with numerical estimations, we need to specify the
input parameters. The effective Wilson coefficient $c_7^\gamma$
including next-to-leading part is $-0.31$ at $\mu=m_b=4.2$ GeV
\cite{Lee}. The effective parameter $a_7^\gamma$ for $B\to
K^*\gamma$ has been calculated in the framework of QCD
factorization to be
$|a_7^\gamma|^2=0.165^{+0.018}_{-0.017}$~\cite{QCDfacBFS} at
$\mu=\hat m_b$ and $a_7^\gamma=-0.4072-0.0256i$~\cite{QCDfacBB} at
$\mu=m_b$ (see also Table 3 of \cite{Ali}). These effective
parameters are larger than the Wilson coefficient $c_7^\gamma$.
For the tensor form factors $T_{1,2}(0)$ with the relation
$T_2(0)=T_1(0)/2$, we use the covariant light-front model result,
namely, $T_1(0)=0.24$ \cite{CC04}.
For the CKM matrix elements, we use the Wolfenstein parameters
$A=0.825$, $\lambda=0.22622$, $\bar \rho=0.207$ and $\bar
\eta=0.340$ \cite{CKMfitter}. For the baryonic form factors, we
will follow \cite{CT96} to apply the nonrelativistic quark model
to evaluate the weak current-induced baryon form factors at zero
recoil in the rest frame of the heavy parent baryon, where the
quark model is most trustworthy. This quark model approach has the
merit that it is applicable to heavy-to-heavy and heavy-to-light
baryonic transitions at maximum $q^2$.  Following \cite{Cheng97}
we have
 \be \label{Lambdabp}
&&
f_1^{\Lambda_b\Lambda}(q^2_m)=g_1^{\Lambda_b\Lambda}(q^2_m)=0.64,
 \quad
 f_2^{\Lambda_b\Lambda}(q^2_m)=g_3^{\Lambda_b\Lambda}(q^2_m)=-0.31,
 \non \\
&&
f_3^{\Lambda_b\Lambda}(q^2_m)=g_2^{\Lambda_b\Lambda}(q^2_m)=-0.10,
 \en
for $\Lambda_b-\Lambda$ transition at zero recoil
$q_m^2=(m_{\Lambda_b}-m_\Lambda)^2$. Since the calculation for the
$q^2$ dependence of form factors is beyond the scope of the
non-relativistic quark model, we will follow the conventional
practice to assume a pole dominance for the form-factor $q^2$
behavior:
 \be
 f(q^2)=f(q^2_m)\left({1-q^2_m/m^2_V\over 1-q^2/m_V^2} \right)^n\,,\qquad
 g(q^2)=g(q^2_m)
\left({1-q^2_m/m^2_A\over 1-q^2/m_A^2} \right)^n\,,
 \en
where $m_V$ ($m_A$) is the pole mass of the vector (axial-vector)
meson with the same quantum number as the current under
consideration. The monopole ($n=1$) momentum dependence is favored
by the pQCD picture. Considering the asymptotic
$t=(p_{\Lambda_b}-p_\Lambda)^2$ limit, two gluons are needed to
distribute the large momentum transfer released from the $b\to s$
transition. Consequently, as $t\to\infty$, we have
$f^{\Lambda_b\Lambda}(t)$ and $g^{\Lambda_b\Lambda}(t)\to 1/t^2$
according to the pQCD counting rule \cite{Brodsky} which gives
rise to the leading power in the large-$t$ fall-off of the form
factor by counting the number of gluon exchanges necessary to
distribute the large momentum transfer to all constituents.

As for the vacuum to $\Lambda \bar p$ vector form factors
$f_i^{\Lambda p}(q^2)$, they can be related to the nucleon's
electromagnetic form factors $F_i^N$ by the relations
 \be
 f^{\Lambda p}_1(t)=-\sqrt{3\over 2}\,F_1^p(t), \qquad\quad
 f^{\Lambda p}_2(t)=-\sqrt{3\over 2}\,F_2^p(t).
 \en
where
 \be
 \la N(p_1)\ov N(p_2)|J_\mu^{\rm em}|0\ra=\bar
 u_N(p_1)\Big[F_1^N(q^2)\gamma_\mu+i{F_2^N(q^2)\over 2m_N}\sigma_{\mu\nu}q^\nu
 \Big]v_{\bar N}(p_2).
 \en
The experimental data are customarily described in terms of the
electric and magnetic Sachs form factors $G_E^N(t)$ and $G_M^N(t)$
which are related to $F_1^N$ and $F_2^N$ via
 \be
 G_E^{p,n}(t)=F_1^{p,n}(t)+{t\over 4m_N^2}F_2^{p,n}(t), \qquad G_M^{p,n}(t)
 =F_1^{p,n}(t)+ F_2^{p,n}(t).
 \en
A phenomenological fit to the experimental data of nucleon form
factors has been carried out in \cite{CHT1} using the following
parametrization:
 \be \label{eq:GMN}
 |G_M^p(t)| &=& \left({x_1\over t^2}+{x_2\over t^3}+{x_3\over t^4}
 +{x_4\over t^5}+{x_5\over t^6}\right)\left[\ln{t\over
 Q_0^2}\right]^{-\gamma},  \non \\
|G_M^n(t)| &=& \left({y_1\over t^2}+{y_2\over
t^3}\right)\left[\ln{t\over Q_0^2}\right]^{-\gamma},
 \en
where $Q_0=\Lambda_{\rm QCD}\approx$ 300 MeV and $\gamma=2+{4\over
3\beta}=2.148$ with $\beta$ being the $\beta$-function of QCD to
one loop. The best fit values of $x_i$ and $y_i$ extracted from
the nucleon data can be found in \cite{CHT2}.

The coupling $g^{\Lambda_b\to B^-p}$ can be fixed from the
measurement of $B^-\to\Sigma_c^0\bar p$ which receives dominant
pole contributions from the intermediate $\Lambda_b^{0(*)}$ and
$\Sigma_b^{0(*)}$ states in the pole model. The expression of the
$B^-\to\Sigma_c^0\bar p$ decay amplitude within the pole model can
be found in Eq. (2.13) of \cite{CY03}. The measured branching
ratio $\B(B^-\to\Sigma_c^0\bar
p)=(3.67^{+0.74}_{-0.66}\pm0.36\pm0.95)\times 10^{-5}$ implies
$|g^{\Lambda_b\to B^-p}|\sim 6$ \cite{Belle:Lamcppi}. As stressed
in \cite{Belle:Lamcppi}, the sign of this strong coupling is fixed
to be negative from other baryonic $B$ decay processes.

\begin{figure}[t]
\vspace{0cm} \centerline{
            {\epsfxsize2.9in \epsffile{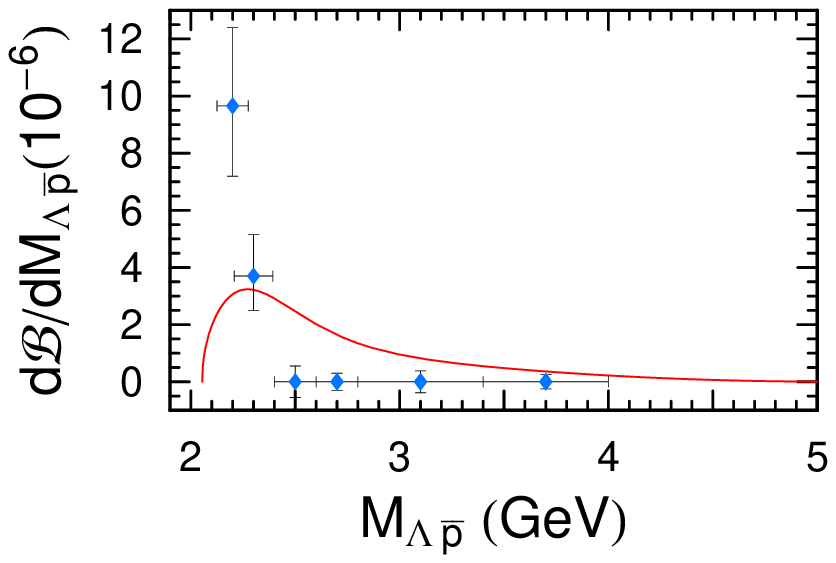}}
            {\epsfxsize3.0in \epsffile{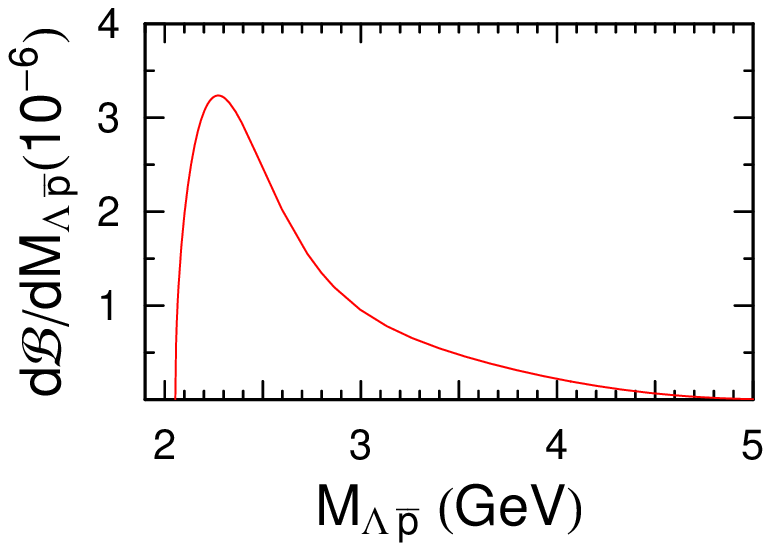}}}
 \centerline{\,\,\,\,\,(a)\hspace{7.2cm}(b)} \vskip0.2cm
\caption{(a) $\Lambda\bar p$ invariant mass distribution for the
decay $B^-\to\Lambda\bar p\gamma$ where the experimental data are
taken from \cite{Belle:Lampgam}, and (b) the same distribution
with amplified vertical scale.} \label{fig:Lampgamma-spect}
\end{figure}

With all the ingredients at hand, we are ready to compute the
decay rate of $B^-\to\Lambda\bar p\gamma$. And the result is shown
in Table I. It is clear that baryon and meson pole contributions
are comparable and interfere constructively. The $\Lambda\bar p$
invariant mass spectrum is depicted in Fig.
\ref{fig:Lampgamma-spect} where the threshold effect, namely, the
peaking branching fraction near the threshold area, is clearly
seen experimentally. \footnote{The data of $d\B/M_{\Lambda\bar p}$
(in units of $10^{-6}({\rm GeV})^{-1}$) are taken from Table I of
\cite{Belle:Lampgam}. For the lower bin $M_{\Lambda \bar p}<2.2$
GeV, $d\B/M_{\Lambda\bar
p}=(1.41^{+0.40}_{-0.36})/(2.2-m_\Lambda-m_p)$.}
Theoretically, the low-mass enhancement effect is closely linked
to the behavior of the baryon form factors in the meson pole
amplitude; the form factors $f_{1,2}^{\Lambda p}(t)$ expressed in
terms of a power series of the inverse of the baryon invariant
mass squared $t=M_{\Lambda \bar p}^2$ will fall off fast with
$M_{\Lambda\bar p}$ [see Eq. (\ref{eq:GMN})]. The threshold
peaking effect will be affected by the presence of the baryon pole
contribution. Specifically, the peak of the spectrum will be
shifted from $M_{\Lambda \bar p}=2.3$ GeV to 2.7 GeV if the $K^*$
pole diagram is turned off while the baryon pole contribution is
turned on. Note that the experimental spectrum is much more
sharply peaked near threshold than the theoretical expectation.
Even if only the meson pole make contributions and the relevant
form factors fall off with $t$ like $1/t^3$ or $1/t^4$,
theoretically it is difficult to fully reproduce the observed
$\Lambda \bar p$ mass spectrum (see also Fig. 2 of \cite{GH}). As
we shall see shortly, the presence of baryon pole contributions is
necessary as it is responsible for the angular distribution to be
discussed below.

\vskip 0.5cm {\bf 4.}~~We next consider other radiative modes such
as $B^-\to\Sigma^0\bar p\gamma$, $B^-\to \Xi^0\ov\Sigma^-\gamma$
and $B^-\to \Xi^-\bar\Lambda\gamma$. Their baryon poles are
$(\Lambda_b^{(*)},\Sigma_b^{(*)})$, $(\Xi_b^0,\Xi_b^{'0})$ and
$(\Xi_b^-,\Xi_b^{'-})$, respectively, where $\Xi_b$ and $\Xi'_b$
are anti-triplet and sextet bottom baryons, respectively. The
relevant strong couplings relative to $g_{\Lambda_b\to B^-p}$
evaluated using the $^3P_0$ quark-pair-creation model are
summarized in \cite{CYrad} and will not be repeated here. The
general feature is that the anti-triplet bottom baryons
$\Lambda_b$ and $\Xi_b$ have larger couplings than the sextet ones
$\Sigma_b$ and $\Xi_b'$. Form factors for $\Sigma_b-\Sigma$,
$\Xi_b-\Xi$ and $\Xi_b'-\Xi$ transitions are depicted in Table 1
of \cite{CYrad}. As for the $K^*$ pole contribution, we apply
again the $^3P_0$ quark model to obtain
 \be
 g_{\Lambda pK^*}=g_{\Xi^-\Lambda K^*}=-\sqrt{2/3}\,g_{\Xi^0\Sigma K^*}
 =-\sqrt{3}\,g_{\Sigma^0 pK^*}.
 \en

\begin{table}[b]
\caption{Branching ratios and angular asymmetries defined in Eq.
(\ref{eq:asy}) for radiative baryonic $B$ decays.} \label{tab:BR}
\begin{ruledtabular}
\begin{tabular}{l l l l c }
 Mode & Baryon pole & Meson pole & Br(total) & Angular asymmetry \\ \hline
$B^-\to\Lambda\bar p\gamma$ & $7.9\times 10^{-7}$ & $9.5\times
10^{-7}$ & $2.6\times 10^{-6}$ & 0.25 \\
$B^-\to\Sigma^0\bar p\gamma$ & $4.6\times 10^{-9}$ &
$2.5\times 10^{-7}$ & $2.9\times 10^{-7}$ & 0.07 \\
$B^-\to\Xi^0\bar\Sigma^-\gamma$ & $7.5\times 10^{-7}$ & $1.6\times
10^{-7}$ & $5.6\times 10^{-7}$ & 0.43 \\
$B^-\to\Xi^-\bar\Lambda\gamma$ & $1.6\times 10^{-7}$ & $2.4\times
10^{-7}$ & $2.2\times 10^{-7}$ & 0.13 \\
\end{tabular}
\end{ruledtabular}
\end{table}

\begin{figure}[t]
\vspace{0cm} \centerline{
            {\epsfxsize2.9in \epsffile{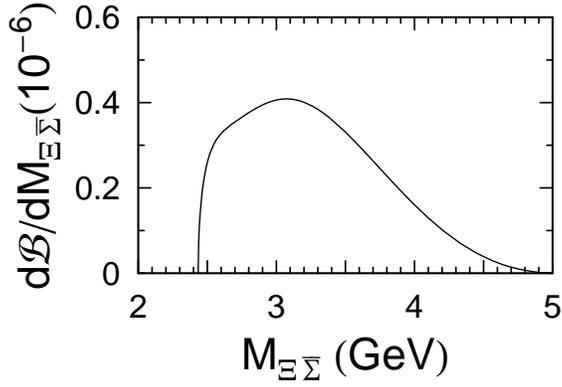}}
            }
\caption{$\Xi\ov \Sigma$ invariant mass distribution for the decay
$B^-\to\Xi^0\bar\Sigma^-\gamma$.} \label{fig:XiSig-spect}
\end{figure}

The predicted branching ratios for $B^-\to\Sigma^0\bar p\gamma,~
\Xi^0\ov\Sigma^-\gamma$ and $\Xi^-\bar\Lambda\gamma$  decays are
summarized in Table 1. Decay rates for the other modes can be
obtained via the relations \cite{CYrad}
 \be
 && \Gamma(B^-\to\Sigma^-\bar n\gamma)= 2\Gamma(B^-\to\Sigma^0\bar
p\gamma)=2\Gamma(\ov B^0\to\Sigma^0\bar
n\gamma)=\Gamma(\ov B^0\to\Sigma^+\bar p\gamma), \non \\
 && \Gamma(\ov B^0\to\Xi^-\bar\Sigma^+\gamma)=2\Gamma(B^-\to\Xi^-\bar\Sigma^0\gamma)=2\Gamma(\ov
 B^0\to\Xi^0\bar\Sigma^0\gamma)=\Gamma(B^-\to\Xi^0\bar\Sigma^-\gamma), \non \\
 && \Gamma(\ov B^0\to\Lambda\bar n\gamma)=\Gamma(
 B^-\to\Lambda\bar p\gamma), \qquad \Gamma(\ov
 B^0\to\Xi^0\bar\Lambda\gamma)=\Gamma(B^-\to\Xi^-\bar\Lambda\gamma).
 \en
It is interesting to notice that the $\Sigma^0\bar p\gamma$ mode,
which was previously argued to be very suppressed due to the
smallness of the strong coupling $g_{\Sigma_b\to B^-p}$, receives
the dominant contribution from the $K^*$ pole diagram. In
contrast, $\Xi^0\bar\Sigma^-\gamma$ is dominated by the baryon
pole contribution. Meson and baryon intermediate state
contributions are comparable in the decays $\Lambda\bar p\gamma$
and $\Xi^-\bar\Lambda\gamma$ except that they interfere
constructively in the former but destructively in the latter. In
short, in addition to the first observation of $\Lambda\bar
p\gamma$, the decay $B^-\to \Xi^0\bar\Sigma^-\gamma$ at the level
of $6\times 10^{-7}$ may be accessible to $B$ factories in the
future. Since this mode is dominated by the baryon pole
contributions, the dibaryon invariant mass spectrum is not sharply
peaked near the threshold as seen in $B^-\to \Lambda\bar p\gamma$.
Rather, it looks like a broad bump with the peak extended away
from the threshold, see Fig. \ref{fig:XiSig-spect}. Hence,
measurements of the invariant mass spectrum for $B^-\to
\Xi^0\bar\Sigma^-\gamma$ will help us understand its underlying
mechanism.

\begin{figure}[t]
\vspace{0cm} \centerline{
            {\epsfxsize3.4in \epsffile{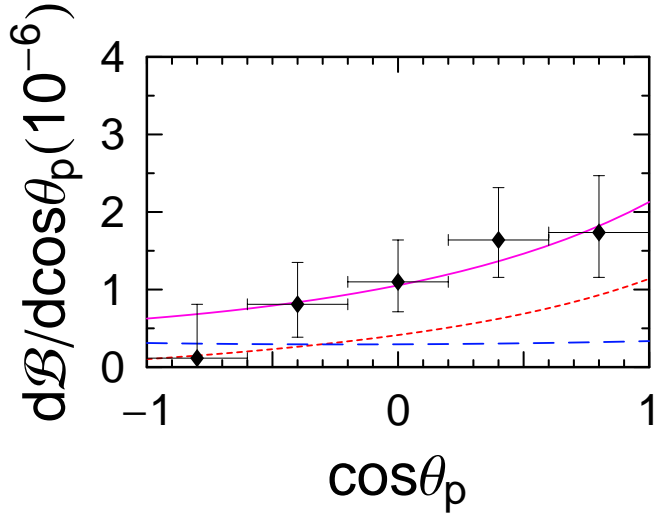}}
            }
\caption{Angular distribution of the antiproton in the baryon pair
system for $B^-\to\Lambda\bar p\gamma$, where $\theta_p$ is the
angle between the antiproton direction and the photon direction in
the $\Lambda \bar p$ rest frame. The dotted, dashed and solid
curves stand for baryon pole, meson pole and total contributions,
respectively. Data are taken from \cite{Belle:Lampgam}.}
\label{fig:Lampgamma-cos}
\end{figure}

\vskip 0.5cm {\bf 5.}~~In addition to the threshold enhancement
effect observed in the differential branching fraction of $\Lambda
\bar p\gamma$, Belle has also measured the angular distribution of
the antiproton in the $\Lambda\bar p$ system (Fig.
\ref{fig:Lampgamma-cos}), where $\theta_p$ is the angle between
the antiproton direction and the photon direction in the $\Lambda
\bar p$ rest frame. It is clear that the $\Lambda$ tends to emerge
opposite the direction of the photon. Defining the angular
asymmetry as
 \be \label{eq:asy}
 A={N_+-N_-\over N_++N_-},
 \en
where $N_+$ and $N_-$ are the events with $\cos\theta_p>0$ and
$\cos\theta_p<0$, respectively, Belle found
$A=0.36^{+0.23}_{-0.20}$ for $B^-\to\Lambda\bar p\gamma$
\cite{Belle:Lampgam}.

The strong correlation of the $\gamma$ to the $\bar p$ in
$B^-\to\Lambda\bar p\gamma$ can be inferred from the baryon pole
diagram in Fig. 1. The denominator of the $\Lambda_b$ pole
amplitude [see Eq. (\ref{eq:Lampole})] reads
 \be
 m_{\Lambda_b}^2-(p_\Lambda+k)^2=m_{\Lambda_b}^2-m_\Lambda^2-2(E_\Lambda
 E_\gamma+p_{\rm c.m.}E_\gamma\cos\theta_p)
 \en
in the $\Lambda\bar p$ rest frame, where $p_{\rm c.m.}$ is the
c.m. momentum. It is clear that the decay becomes prominent as
$\cos\theta_p\to 1$, i.e., when the antiproton is parallel to the
photon. The meson pole diagram is responsible for low-mass
enhancement and does not show a preference for the correlation
between the baryon pair and the photon (see the dashed curve in
Fig. \ref{fig:Lampgamma-cos}). Our prediction $A=0.30$ (see Table
\ref{tab:BR}) is consistent with experiment. Intuitively, the
observation that $\gamma$ is correlated more strongly to the $\bar
p$ than to the $\Lambda$ also can be understood in the following
manner. Since in the $B$ rest frame
 \be
 M_{\Lambda \bar p}^2=m_\Lambda^2+m_p^2+2(E_\Lambda E_{\bar
 p}-|\vec{p}_\Lambda||\vec{p}_{\bar p}|\cos\theta_{\Lambda\bar
 p}),
 \en
threshold enhancement implies that the baryon pair $\Lambda$ and
$\bar p$ tend to move collinearly in the $B$ rest frame, i.e.
$\theta_{\Lambda\bar p}\to 0$. From Fig. 1 we see that the
$\Lambda$ is moving faster than $\bar p$ as the former picks up an
energetic $s$ quark from the $b$ decay.  When the system is
boosted to the $\Lambda$ rest frame, $\bar p$ and $\gamma$ are
moving collinearly away from the $\Lambda$.

We next turn to the $\Sigma^0\bar p\gamma$ and $\Xi^0\bar
\Sigma^-\gamma$ modes. Since they are dominated by meson and
baryon pole contributions, respectively, the angular asymmetry is
expected to be very large ($\sim 45\%$) for the former and very
small $(\sim 8\%$) for the latter.

In \cite{GH}, the amplitude of $B^-\to\Lambda\bar p\gamma$ is
expressed in terms of two 3-body matrix elements $\la\Lambda\bar
p|\bar s\gamma_\mu(1-\gamma_5)b|B^-\ra$ and $\la\Lambda\bar p|\bar
s(1+\gamma_5)b|B^-\ra$. These matrix elements can be parameterized
in terms of several unknown form factors. If the form factors are
expanded as a power series of the inverse of the baryon invariant
mass $(1/M_{\B_1\bar \B_2}^2)^n$, then the angular distribution
curve would look like a parabola opening downward as the photon
has no preference in correlation with the $\Lambda$ or the $\bar
p$ \cite{SYTsai}, in sharp contrast to the data shown in Fig.
\ref{fig:Lampgamma-cos}. Phenomenologically, this means that the
$B\to\B_1\ov\B_2$ transition form factors should not depend solely
on the baryon invariant mass; the invariant mass of the photon and
one of the baryons should be incorporated in the form-factor
momentum dependence \cite{SYTsai}. This difficulty is resolved in
the pole model as the invariant mass of the $\gamma$ and the
$\Lambda$ is included in the propagator of the meson pole
contribution to $B^-\to\Lambda\bar p\gamma$. The issues for the
angular distributions in baryonic $B$ decays $B\to\B_1\ov \B_2M$
and for the momentum dependence of $B\to\B_1\ov\B_2$  transition
form factors will be studied elsewhere.

\vskip 0.5cm {\bf 6.}~~ We have re-examined the weak radiative
baryonic $B$ decays $\ov B\to\B_1\ov \B_2\gamma$ mediated by the
electromagnetic penguin process $b\to s\gamma$ within the
framework of the pole model.  The meson pole contribution that has
been neglected before is taken into account in this work. Our
conclusions are: (1) The intermediate $K^*$ contribution dominates
in the $\Sigma\bar p\gamma$ mode and is comparable to the baryon
pole effect in $\Lambda\bar p\gamma$ and $\Xi\bar\Lambda\gamma$
modes. (2) The branching ratios for $B^-\to\Lambda\bar p\gamma$
and $B^-\to\Xi^0\bar\Sigma^-\gamma$ are of order $2.6\times
10^{-6}$ and $6\times 10^{-7}$, respectively. (3) The meson pole
contribution accounts for the threshold enhancement effect
occurred in the dibaryon invariant mass spectrum. (4) The baryon
pole diagrams imply that the antibaryon tends to emerge in the
direction of the photon in the dibaryon rest frame. Our predicted
angular asymmetry $A=0.25$ agrees with experiment for
$B^-\to\Lambda\bar p\gamma$. Measurements of the correlation of
the photon with the baryon allow us to discriminate between
different models for describing the radiative baryonic $B$ decays.
(5) For the decay $B\to\Xi\bar\Sigma\gamma$, a large correlation
of the photon to the $\bar\Sigma$ and a broad bump in the dibaryon
mass spectrum are predicted.

\vskip 2.5cm \acknowledgments We are grateful to Min-ru Wang for
discussion. This research was supported in part by the National
Science Council of R.O.C. under Grant Nos. NSC94-2112-M-001-023
and NSC94-2112-M-033-001.

\pagebreak


\end{document}